\documentclass[twocolumn,10pt,amssymb,aps,showpacs,prd]{revtex4-1}

\usepackage{graphicx}
\usepackage{amsmath}
\usepackage{bm}
\usepackage{natbib}
\begin{document}
\title{\bf Gravitational collapse with equation of state} 
%%%
\author{Sanjay Sarwe$^{1,}$}\email{Electronics address:
sbsarwe@gmail.com} \ \author{R. V. Saraykar$^{2,}$}\email{Electronics 
address:ravindra.saraykar@gmail.com} \ \author{Pankaj S. 
Joshi$^{3,}$}   \email{Electronics address:psj@tifr.res.in}

\affiliation{{$^{1}$ Department of Mathematics, 
S. F. S. College, Seminary Hill, Nagpur-440 006, India}\\
{$^{2}$ Department of Mathematics, RTM Nagpur University 
Campus, Nagpur-440 033, India}\\
{$^{3}$ Tata Institute of Fundamental Research, 
Homi Bhabha Road, Colaba, Mumbai-400 005, India}}

\begin{abstract} 
To get deeper insight about gravitational collapse 
of a perfect fluid with a linear isentropic equation of state 
$p = k \rho$, we consider a model which includes a family 
of solutions to Einstein equations with equation of state. 
The collapse evolves from a regular initial 
data and the positivity of energy conditions, and other physically 
relevant regularity conditions are satisfied. 
The final fate of collapse is
analyzed in terms of the formation of black holes and naked 
singularities. 
This gives useful insights into the end states 
of collapse with a physically  relevant equation 
of state in the light of cosmic censorship hypothesis.   
\end{abstract}  


%Keywords: {Type I matter field, gravitational collapse,
%equation of state, central-singularity}
\maketitle

\section{Introduction}
The cosmic censorship conjecture (CCC) articulated 
by Penrose 
\cite{rp}
is fundamental to many aspects of theoretical and astrophysical 
applications of black hole physics today. Despite many attempts
over past decades no theoretical proof or even any 
satisfactory mathematical formulation of CCC is available 
as of today in the case of dynamical gravitational collapse.
 In the mean time, many authors have studied 
 spherical gravitational collapse of a massive matter 
cloud within the framework of general relativity. As the 
nuclear fuel of a massive star gets exhausted, it loses its 
equilibrium and gravity becomes the central dominant force 
which lends the star to its perpetual collapse. 
The gravitational collapse studies then show that the 
collapse end state is either a black hole (BH) or a naked 
singularity (NS), depending on the nature of the initial 
data from which the collapse evolves, arising from a regular 
initial state to the final super dense state. 
\cite{jg,jr1,r1,op,dj2,wl,ct,rw}

The works such as above analyze physically relevant
general matter fields that include most of the known
physical forms of matter like dust, perfect fluids, etc. 
Wide classes of solutions to Einstein equations 
are shown in these cases to exist where the collapse end state 
is a black hole or naked singularity depending on the nature of 
initial data and the collapse evolutions allowed by the 
Einstein equations subject to regularity and physical 
reasonability conditions. 

An equation of state for the matter is, however, not always 
assumed in many of these investigations. As such, the physical 
characteristics of the matter field constituting the star 
are described by an equation of state relating the density 
and pressures, and therefore it is important to know if 
the naked singularities would still form when a suitable 
equation of state is assumed for the collapsing cloud. 
But it is to be noted that such an equation of state is 
not precisely known for a collapsing massive star which 
will be assuming super dense states of matter closer to 
the final later stages of the collapse, where the physical 
region has ultra-high densities, energies and pressures. 
It is far from clear as of now what would be a physically 
realistic equation of state that would describe such an 
ultra-high density region of collapse, and also whether 
the equation of state would remain unchanged or it would 
actually evolve and change as the collapse develops.

Nevertheless it is important to understand the final fate
of a gravitational collapse when a suitable equation of state 
is assumed, and to know if the naked singularities still
form as collapse final states. One can thus choose the equation of 
state to be linear isentropic or polytropic to describe the 
collapse of a massive star, right after its departure from the 
equilibrium configuration where gravity was balanced by the 
nuclear reactions taking place at the core region of the 
star. 

The gravitational collapse of a perfect fluid with a linear 
equation of state is of interest from both theoretical as well
as numerical relativity perspectives. 
Our motivation also comes from certain other questions 
such as, what if the value of $k$
increases in the range $0 \leq k \leq 1$ when a naked
singularity appears as collapse final state for a given 
spacetime dimension and for a critical positive tangent
to the singularity curve as to be discussed below. 
In such a case will the singularity sustain its 
nature, or a formation of event horizon shall precede the 
formation of singularity to preserve the CCC? We believe
that answer to these and similar issues would be important 
to understand better the physical conditions for collapse
and the role of an equation of state towards providing 
a suitable mathematical and physical formulation 
of CCC.

We therefore consider here a linear isentropic equation 
of state, $p = k \rho,\; \; 0  \leq k \leq 1$ in the study of 
spherical gravitational collapse of a perfect fluid. 
Self-similar perfect fluid collapse models with a linear 
equation of state were considered through numerical simulations 
by Ori and Piran
\cite{op}
and analytically by Joshi and Dwivedi 
\cite{dj2}
to show how black holes and naked singularities develop
as collapse final states in this scenario. 
Further, Goswami and Joshi
\cite{rgj} 
studied the case of an isentropic perfect fluid 
with a linear equation of state in four dimensional spacetime 
without the self-similarity assumption, wherein they
showed that the occurrence of BH and NS evolving from regular
initial data depends on the choice of rest of the free 
functions available. It was also proved that in a general 
$N$-dimensional dust collapse, the occurrence of NS can be 
removed when one goes to a higher dimensional spacetime, 
thus restoring CCC. 
\cite{rgpj-1}
Dadhich 
{\it et al}  \cite{DSD} 
studied spherically symmetric collapse of  
a fluid with a non-vanishing radial pressure in higher 
dimensional spacetimes with an equation of state $p_{r} =k \rho$ 
using the construction of the Joshi-Dwivedi root equation 
that governs the nature (BH versus NS) of the central singularity. 
Also other cases of collapse with pressure have been studied 
by various authors to understand the collapse
final fate of collapse
 \cite{JhHaGiHa}.

Inspite of all these works,  we still do not know 
the role of general pressures during the end stages of collapse. In order to have an insight, the study of perfect fluid collapse has been carried out mostly under some simplifying assumptions and restrictions. A general formalism for perfect fluids
with a physically relevant equation of state is still lacking due to the intrinsic
difficulties posed by Einstein equations. In gravitational collapse perfect fluids models appear as a natural choice and they are commonly used
since these are the models that describe gravitating stars in
equilibrium and since it can be shown that near the center of the cloud regularity
implies that matter must behave like a perfect fluid
  \cite{DPJRVS}.

Our aim here is to study a model which
includes a class of collapse 
models for the perfect fluid case with a linear equation of state.
We analyze the same for the collapse end states in terms 
of formation of black holes and naked singularities. We analyze 
here the marginally bound case for the sake of simplicity 
and transparency and the results show that BH and NS form 
for a wide range of values of the parameter $k$ in the 
equation of state $p = k \rho$.   

The plan of work is as follows: In section II, 
we study Einstein field equations for spherically symmetric 
metric. We obtain the function $M(r,v)$
as a class of solutions to the constraint equation that 
enters through dynamical equations for $p=k \rho$. 
The dynamics of collapse for occurrence of NS/BH 
phases due to the effect of regular initial data is studied 
in the subsection III. To illustrate the end state of collapse, a special case
$g(v)=0$ at $r=0$ is considered and it is studied in subsection III.1.
The nature of radial null geodesic emanating from the singularity 
is discussed in section IV. The occurrence of NS/BH phases is investigated for increasing 
value of the parameter $k$, and illustration
of its behavior in dust collapse is provided in subsection IV.1 and that of radiation collapse 
in subsection IV.2.
The implication of the function $M(r,v)$ on singularity curve is  
investigated, which corroborate the study of apparent horizon in section V. 
The conclusions are specified in section VI.

\section{Collapse with a linear equation of state}
The general spherically symmetric metric describing 
spacetime geometry of a collapsing cloud can be described in
comoving coordinates \ $(t, r, \theta, \phi)$ 
by the metric,
\begin{equation}
 ds^2 = - e^{2 \nu(t,r)}dt^2 +  e^{2 \psi(t,r)}dr^2 + R^2(t,r) d\Omega^2 \label {Dm01}
\end{equation}
where 
\[ d\Omega^2= d \theta^2 + \sin^2(\theta) d \phi^2 \] 
is the
metric on a two-sphere.
We aim here at the question, whether the choice of
an equation of state $ p = k \rho$, $k \in [0,1]$ contributes 
through the parameter $k$ in the development of BH/NS phases. 
The stress-energy tensor for the type I matter fields,
and in particular for perfect fluids, is in a diagonal form 
in a comoving coordinate system and is given by 
\cite{HE}, 
\[
 T^{t}_{t}  = -\rho, \; T^{r}_{r}  =
 T^{\theta}_{\theta} = T^{\phi}_{\phi}=\; p \;.
\] 
The quantities $\rho$ and $p$ are the energy
density and pressure respectively. We take the matter field to
satisfy the weak energy condition which implies ${{\rho} \ {\geq}}0$;
${\rho+p \ {\geq}} 0$. The linear equation of state for a 
perfect fluid is,
\begin{equation}
p(t,r) = k \ \rho(t,r) \ \text{where} \  k \in [0, 1]\; . \label{Des04}
\end{equation}
The Einstein field equations for the metric 
(\ref {Dm01}) are derived as 
\cite{rgj}
\begin{equation}
 \rho = \frac{{F}'}{ R^{2} R'} \;  \;
 = - \frac{1} {k} \frac{\dot{{F}}}{R^{2} \dot{R}} \label {Dsfe05} 
\end{equation}
\begin{equation}
 \nu \;' = - \frac{k}{k + 1} [ \ln(\rho)]' \label {Dsfe06} 
\end{equation}
\begin{equation}
 R' \dot{G} - 2 \dot{R} G \ {\nu}' = 0  \label {Dsfe07} 
\end{equation}
\begin{equation}
 G - H = 1 - \frac{{F}}{R} \; \; . \label {Dsfe08}
\end{equation}
Note that `prime'  and `dot' denote differentiation with respect to $r$ and $t$.
Here $F = F(t, r)$ is an arbitrary function, and 
has an interpretation of the mass function for the cloud. 
It gives the total mass in a shell of comoving radius $r$ 
on any spacelike hypersurface $t = const.$ The energy conditions 
impose the restriction on $F$, namely that $F(t,r) \geq 0$. 
In order to preserve the regularity at all the epochs, 
we have $F(t, 0) = 0$, that is, the mass function has to 
vanish at the center of the cloud. Since we are considering 
collapse, we have $\dot{R} < 0$, {\it i.e.} the physical radius 
$R$ of the collapsing cloud keeps decreasing in time and 
ultimately it reaches $R = 0$, which denotes the spacetime 
singularity where all the matter shells collapse to a zero 
physical radius. The functions $G$ and $H$ are defined as 
$G(t, r) = e^{-2 \psi} {R'}^2$ and $H(t, r) = e^{-2 \nu} \dot{R}^2$.

Let us define a new function  $v(t, r)$ by
$v(t, r)= R/r$,  
and now we use the scaling independence of the comoving
coordinate $r$ to write
\begin{equation}
R(t, r) = r \; v(t, r) \; .\label{Df09}
\end{equation}
This gives $v(t_{i}, r) = 1, v(t_{s}(r), r) = 0$ and
the collapse condition is now written as $\dot{v} < 0$.
The time $t = t_{s}(r)$ corresponds to the shell-focusing 
singularity at $R = 0$, where all the matter shells collapse to a
vanishing physical radius. The introduction of the parameter 
$v$ as above allows us to distinguish the genuine spacetime 
singularity from the regular center at $r=0$, with $v = 1$ 
at the initial epoch including the center $r = 0$, and $v$ 
then decreases monotonically with time as collapse progresses 
to the value $v = 0$ at the singularity $R = 0$. 
At the regular center the mass function $F(t,r)$ behaves 
suitably so that the density remains finite
and regular there at all times till the occurrence 
of singular epoch. 
From the Einstein equation for density, we see that 
the mass function for the cloud can be written in general 
as,
\begin{equation}
{F}(t,r) = r^{3} M(r,v) \label {Dsm12}
\end{equation}
where $M(r,v)$ is regular and continuously twice
differentiable. Using equation (\ref {Dsm12}) in equations (\ref
{Dsfe05}), we obtain
\begin{equation}
\rho =  \frac{3 M + r [ M_{, r} + M_{, v}  \ v']}{ v^{2} ( v + r v')}  
  =   - \frac{ M_{, v}} { k \ v^{2}}.  \label {Dsm13}
\end{equation}
Then as $ v \rightarrow 0 \; , \rho \rightarrow \infty $ and
 $p \rightarrow \infty $, {\it  i.e.} both the density and pressure 
blow up at the singularity. 
We rearrange equation (\ref{Dsm13}) as follows,
\begin{equation}
 k r {M}_{,\ r} +[(k + 1) r v' + v] {M}_{,\ v} = - 3 k M \; . \label {Dsv14.1}
\end{equation}
The general solution of equation (\ref {Dsv14.1}) represents 
many classes of solutions but only
those classes are to be considered which satisfy
the energy conditions, which are regular and those which
give $\rho \rightarrow \infty$ as $v \rightarrow 0$. 
This means that the energy conditions and
equation of state $ p = k \rho$ isolate the class of functions
${M}(r,v)$ so that the mass function ${F}(t,r)$,
the metric function $\nu (t,r)$ and other concerned functions 
of $v$ evolve as the collapse begins according to the Type I
field equations. 

To obtain such a class of solutions, 
 we consider here the ansatz, 
\begin{equation}
\frac{r v'} {v} = g(v) \label{CON01}
\end{equation}
due to which the equation (\ref {Dsv14.1}) takes the form
\begin{equation}
 k r {M}_{,\ r} +[(k + 1)  v g(v) + v] {M}_{,\ v} = - 3 k M \label {Dsv14A.1}
\end{equation}
wherein in order to analyze the final outcome of collapse, we set $r$ and $v$ 
as independent coordinates by performing a transformation
from $(r, t)$ to $(r, v)$, thus considering $t = t(r, v)$. Now, above equation 
has a general solution of the form,
\begin{eqnarray}
&& M(r,v) = m_{o} \; f{( x )} \; e^{- 3 Z(v)}\; \;\text{where} \;\; x= re^{-Z(v)} \label{SoD15} \\
&& \text{and} \; Z(v)  = \int_{1}^{v} \frac {k} {v[(k+1) g(v) + 1]} dv \label{SoD15.1}
\end{eqnarray}
where $m_{o}$ is a positive constant.
We note that $ g(1) =0 $ and assume that 
\[ g(0)= \lim_{ v \rightarrow 0} \frac{r \; v'}{v}= \alpha_{o}  \;  \; \text{exists.} \]
  %%%%%%%%%%%%%%%%%%%%%%%%%%%%%
	
The existence of general solution of equation (\ref{Dsv14A.1}) is one more daunting question
which we verify in the following proposition. \\
Theorem 1:  The general solution of equation (\ref{Dsv14A.1}) exists in domain of 
$v \in (0,1]$ and $ 0 \leq r \leq r_{b}$
if  $f(x)$ and $g(v)$ are continuously differentiable functions in the domain such that 
the $ \lim_ { v \rightarrow 0} g(v) $  exists. \\
Proof:  The assumed properties of $g(v)$ allow us to infer that the integrand of  equation (\ref{SoD15.1}) 
is continuous and therefore $Z(v)$ is integrable, and so $Z_{,v}$ exists.  Hence the shorthand
$x = r e^{-Z(v)}$ is well defined in the domain. \\
The general solution prescribed in equation (\ref{SoD15}) gives partial derivatives 
\begin{eqnarray*}
&& M_{,r}= m_{o} e^{-4 Z(v)} f '(x), \nonumber       \\
&& M_{,v}= - \frac{ m_{o} k e^{-3Z(v)} [ 3 f(x) + x f '(x)] } {v [ (k+1) g(v) + 1] } 
\end{eqnarray*}
and since the function $ f '(x)$ exists, theses partial derivatives exist in the domain 
$v \in (0,1]$ and $ 0 \leq r \leq r_{b}$.  Hence the proof. $\diamond$ \\
In $(t, r)$ coordinates, the mass function $F(t, r)$ using  equations (\ref{SoD15}, \ref{SoD15.1})
satisfies the field equation (\ref{Dsfe05}), and hence we have a general solution of field equation (\ref{Dsfe05}).
%%%%%%%%%%%%%%%%
%%%%%%%%%%%%

The introduction of the equation
(\ref{CON01}) demands its compatibility with other 
field equations or their subsequent 
equation (\ref{RD19}), which is discussed in the Appendix I. 
 
 The density profile 
for this class of models then takes the form,
\begin{equation}
\rho(r,v) = \frac{ m_{o} e^{- 3 Z(v)}\; [ 3 f(x) + x f'(x)]} 
{v^{3}[(k+1) g(v) + 1] }.  \label{Sr16} 
\end{equation}
We note that at the initial epoch, $g(1)=0,Z(1) =0$, 
and the regular density distribution takes the form
\begin{equation}
\rho_{o}(r) = m_{o}\;   [3 f(r) + r f'(r)] . \label{Sr17}
\end{equation}
Such a density profile should be decreasing away from the 
center $r=0$, which is a physically reasonable feature for
the collapsing matter cloud, and since $\rho_{0}(r) \geq 0$, we have
$[3 f(r) + r f'(r)] \geq 0 $. Further, at the epoch 
$v=1$, $\rho_{o}(r)=0$ at some value 
$r= r_{b}$ which is determined from $f'(r)/f(r) = -3/r$, it gives 
$f(r) = const./r^3$.
 Hence we would take such a value of $r=r_{b}$ 
as the boundary of the cloud where the energy density 
is zero, and where the interior is matched to a suitable
exterior metric. 

The energy density at the center of the collapsing cloud takes the form 
\begin{equation}
\rho(0,v) = \frac{ 3 m_{o} e^{- 3 Z(v)}\;  f(0) } 
{v^{3}[(k+1) g(v) + 1] },  \label{Sr17A} 
\end{equation}
and we observe that $\rho$ satisfies the energy condition $\rho \geq 0$ 
at all the epochs.

The gradient of $\rho(r,v)$ is obtained as
\begin{eqnarray}
 \rho_{,r}(r,v)&& = \frac{- m_{o} e^{-3Z} } 
{ v^{3}[(k+1)g(v)+1]^2}  \times  \nonumber \\
&& \Big\{(3 f(x) + x f'(x))   \frac{v'}{v}(k+1)[3(1+g(v)) + v\frac{dg}{dv}]  \nonumber \\
&& - (g(v)+1) e^{-Z} [4 f'(x) + x f''(x) ] \Big\}. \label{DGRAY}
\end{eqnarray}
At the initial epoch $t=t_{i},v=1$, we get $\rho_{,r}=0$ at the center 
$r=0$ when $f'(0)=0$. The radial coordinate $r$ has the range 
$0 \leq r \leq r_{b}$ which defines the interior of the 
compact collapsing object, and wherein the density profile 
is smooth and monotone decreasing away from the center of 
the cloud as one moves towards the boundary of 
the cloud. The above discussion give rise to the following theorem. \\
Theorem 2: Consider the equation of state $p = k \rho$, the mass profile $F(t, r) = r^3 M(r, v)$
 and the density profile as in equation (\ref{Dsm13}). Now, if
$r v '(t, r) = v(t, r) g(v(t, r))$ is introduced as an additional equation in the set of field equations
then the initial data of mass function, and thereof density
is non-singular at the initial epoch $t = t_{i}$. \\
Proof:  The mass and density profiles, and their derivatives at $t = t_{i}$ take the forms as follows:
\begin{eqnarray}
 F(t_{i}, r) &&= r^3m_{o} f(r) \nonumber \\
 F'(t_{i}, r) &&= m_{o}\;r^2 [3 f(r) + r f'(r) ] \nonumber \\
 \dot{F}(t_{i}, r ) &&= - k \; m_{o} \dot{v}(t_{i}, r)\; r^3 [3 f(r) + r f'(r) ]  \nonumber \\
 \rho(t_{i},  r) &&= m_{o} [3 f(r) + r f'(r) ]  \nonumber \\
 \rho'(t_{i}, r) &&=  m_{o} [4 f'(r) + r f''(r) ]  \nonumber \\
 \rho''(t_{i}, r) &&= m_{o} [5 f''(r) + r f'''(r) ]  \nonumber \\
 \dot{\rho}(t_{i}, r)&&= - m_{o}  \dot{v}(t_{i},r) \Big[ 4 k r f'(r) + k r^2 f''(r) \nonumber \\
&& +( 3 f(r) + r f'(r) ) (k+1) (3 + \frac{dg}{dv}) \Big] \label{IDMR}
\end {eqnarray}
The regularity conditions as explained earlier at the initial epoch demand $f(r)$ to be a non-decreasing function 
such that $[3 f(r) + r f'(r)] \geq 0 $, and $[4 f'(r) + r f''(r) ] \leq 0$
and further at ($t=t_{i}, r=0)$, we need $f(0)= $ a finite positive number, $f'(0)=0$ and $f''(0)= $ a negative number. 
Thus, the Einstein field equations will have a class of solutions having non-singular
initial data of mass and density profiles,
formed due to the legitimate choice of function $f(x)$ satisfying aforementioned conditions.$\diamond$

The above conditions satisfy for the class of functions such as $f(x) = 1 - x^2/2$, etc. Herein for simplicity,
we have chosen $f(0)= 1$, $f '(0)=0$ and $f ''(0)= -1 $. 
From equation (\ref{SoD15}), we can write
\begin{eqnarray}
&& M_{0}(v)=M(0,v)= m_{o} e^{-3Z}, \; 
M_{1}(v)= \frac{M_{,r}(0,v)} {1 !}=0 \; \nonumber \\
&& \text{and} \;  
M_{2}(v)= \frac{M_{,rr}(0,v)} {2 !}= - \frac{m_{o} e^{-5Z}} 
{2}. \hspace{0.5in}\label{MEQs}
\end{eqnarray}
Here $M_{1}(v)=0$ is in accordance with the requirement that 
the energy density has no cusps at the center and that its first
derivative vanishes at the center.

\section {Dynamics of collapse and the BH/NS phases}
Based on the above, we would now like to determine the 
final states for such a perfect fluid collapsing cloud in terms
of the black holes and naked singularities. As we shall
point out, the key factor that determines this final
outcome is the geometry of the trapped surfaces. If the 
trapped surfaces developed early enough in collapse, well 
before the formation of the singularity, then a black hole 
is the collapse final state. On the other hand, when the 
trapped surface formation is delayed as the collapse evolves,
due to the internal dynamics of the collapsing cloud, 
in that case the singularity is no longer fully covered 
by an event horizon, and we can have light rays and 
material particles escaping away from an arbitrary vicinity 
of the same, thus giving rise to a naked singularity.   

To decide and understand this trapped surface geometry,
we need to determine the behavior of the singularity curve
in the spacetime, which describes how and when different 
collapsing shells arrive at the zero physical radius, 
thus forming the spacetime singularity.

To determine this, 
firstly on integrating equation (\ref {Dsfe06}), we obtain 
the general metric function,
\begin{equation}
\nu(r,t) = - l\; \ln(\rho) + a(t). \label {Dcv14}
\end{equation}
where $ l = \frac{k} {k + 1} $. We can set $a(t) =0$. 
Let $A(r, v)$ be a suitably differentiable function 
defined by $A(r, v)_{,\ v} = {\nu \;'} / {R'}$, and on using 
equations (\ref{CON01}) and (\ref {Dcv14}), it takes the form
\begin{eqnarray}
A(r,v)_{, v} = && - \frac{l \; {\rho}'} { {\rho} R'}  \nonumber \\
= && \frac{-l}{v[(k+1)g(v)+1]} 
\Big\{ e^{-Z}\left[ \frac{4 f'(x) + x f''(x)}{3f(x)+x f'(x)}\right]   \nonumber \\
&&- \frac{v'(k+1) [3(1+g(v)) +v \frac{dg}{dv}]}{v(1+g(v))}         
\Big\}.  
 \label {Decv15}
\end{eqnarray}
Then on integrating equation (\ref{Decv15}), we obtain
\begin{eqnarray}
A(r,v)  
&&=\int_{v}^{1} \frac{-l}{v[(k+1)g(v)+1]} 
\Big\{e^{-Z}\left[ \frac{4 f'(x) + x f''(x)}{3f(x)+x f'(x)}\right]     \nonumber \\
&& - \frac{W(k+1) [3(1+g(v)) +v \frac{dg}{dv}]}{v(1+g(v))} 
\Big\} dv 
\label{Decv16}
\end{eqnarray}
where $v'(t,r)=W(r,v)$. The regularity
at the center of the cloud implies $W \rightarrow 0$ as $r \rightarrow 0$.
We note that if we explicitly find $A(r,v)$, we then have an 
exact solution of the Einstein field equations for the perfect
fluid collapse case. As is known this is of course a complex 
problem, and in the present scenario, the unknown nature of 
the function $g(v)$ forbid such a possibility of 
writing down the exact metric for the full spacetime. 
However, we can still write the metric in the neighborhood 
of the center of the cloud as we point out below. 

Our aim below is to find the BH/NS phases for collapse 
for a given initial data in terms of the function $M(r,v)$.
The mass function $M(r,v)$ satisfies the regularity
condition $M_{,r}(0,v)=0$ and we have $A \approx r$
closer to the center $r=0$.

The function $A$ can be expressed 
as an expansion and it takes the form,
\begin{equation}
A(r, v) = r A_{1}(v)+ r^{2} A_{2}(v) + r^3 A_{3}(v) + 
{\mathcal{O}}(r^{4}), \label{ARDef1}
\end{equation}
wherein
\begin{eqnarray}
 A_{1}(v) &&= \frac{A_{,r}(0,v)} {1!} =  
\frac{5 \;l} {3} \int_{v}^{1} \frac{e^{- 2 Z(v)}} {v} dv ,  \nonumber \\
 A_{2}(v) &&=  \frac{A_{,rr}(0,v)} {2!} = 
- l \; f '''(0) \int_{v}^{1} \frac{e^{- 3 Z(v)}} {v} dv  \nonumber \\
 A_{3}(v) &&=  \frac{A_{,rrr}(0,v)} {3!} = 
\frac{l}{18} 
\int_{v}^{1} \frac{[25 - 7 f ''''(0)] e^{- 4 Z(v)}} {v} dv.
 \label{As17A}
\end{eqnarray}
 While finding the values of $A_{1}$ etc, we restrict our analysis to constant $v$ surfaces,
so we can put $W$ and its derivatives to be zero in $ A_{1}, A_{2}, A_{3}$.
On the other hand if we approach the singularity along a
generic curve we cannot neglect the terms in $W$.

Also, the use of definition of $A(r,v)$ in equation (\ref
{Dsfe07}) yields
\begin{equation}
G(t,r) = \textbf{d}(r) \; e^{2rA} \label {Dg18}
\end{equation}
where $ \textbf{d}(r) $ is another arbitrary continuously
differentiable function of $r$. Following comparison with the dust
collapse models, we can write
\begin{equation}
\textbf{d}(r) = 1 + r^2 \; \textbf{b}(r)  \label {Db17}
\end{equation}
where the function $\textbf{b}(r)$ basically characterizes 
the energy distribution for the collapsing shells.
For the sake of simplicity and transparency of consideration
and for a better understanding of the model, 
we consider here only the `marginally bound case',
which corresponds to $\textbf{b}(r)=0$.

As we see now, having supplied an explicit 
function $M(r,v)$ which is a solution to the first two 
Einstein equations, that determines $\rho(r,v)$,
and therefore the metric function $\nu(r,v)$ is fully 
determined as above. The function $A(r,v)$ is 
also determined as above and is now a known function, 
and therefore $G(r,v)$ is also determined. Finally, the 
metric function $R$ is determined as we shall mention 
below.  We can write the metric in the 
neighborhood of the center 
$r=0$ of the cloud as,
\begin{equation}
 ds^2 = - \rho^{-2\; l}dt^2 +  \frac{R'^{2}}{ e^{2rA}} dr^2 
+ R^2(t,r) d\Omega^2  \; .\label {Dms01}
\end{equation}

In order to determine the nature of the singularity
curve $t_s(r)$ which corresponds to the physical singularity 
at $R=0$ or $v=0$, 
the field equation (\ref{Dsfe08}) can be written in the form
\begin{equation}
\dot{R}^2 = e^{2 \nu} \left[\frac{F}{R} + G -1 \right]  \; . \label{RD18}
\end{equation}

In the study of final stages of collapse of a dense cloud, and the formation of BH
and NS, we must have the initial configuration to be not trapped.
This allows for the formation of trapped surfaces during collapse and therefore
we must have
\[\frac{F(t_{i},r)} {R(t_{i},r) } = r^2 M_{o} f(r) < 1
\]
from which we observe, how the choice of the initial matter configuration $M(r,1)$ through $f(r)$ is related
to the initial boundary of the collapsing cloud. Some restrictions on the choices of
the radial boundary must be made in order not to have trapped surfaces at the
initial time. This condition is reflected on the initial configuration and gives some constraints on the initial
velocity of the infalling shells. So, to avoid trapped surfaces at the initial epoch the velocity of the
infalling shells must satisfy \ \
$|\dot{R} | > \sqrt{ |d(r)|} \ e^{r A + \nu} $ and it shows 
 that the initial velocity of the cloud must always be positive and that the
case of equilibrium configuration where $\dot{R} = 0$ can be taken only 
at the static boundary of the star where pressure is zero.
\cite{DPJRVS}

Using equations (\ref{Dsm12}) and (\ref{Dg18}), equation (\ref{RD18})  
can be expressed as,
\begin{equation}
\dot{v} = - e^{\nu} \sqrt{\frac{M(r,v)}{v} 
+\frac{ e^{2rA} -1}{r^2} } \label{RD19}
\end{equation}
where the negative sign is chosen since for collapse
we must have $ \dot{v} < 0 $.
Integrating the above equation, we have
\begin{equation}
 t(r,v) = t_{i} + \int_{v}^{1}  \frac{ e^{- \nu} dv}
 { \sqrt{\frac{M(r,v)}{v} +\frac{ e^{2rA} -1}{r^2} }} \; \; .
\hspace{.4in} \label {Dsi21}
\end{equation}
In above equation, the variable $r$ is treated as a constant.
Regularity ensures that, in general, $t(r, v)$ is 
at least $C^{2}$ near the singularity and therefore can be 
expanded around the center as,
\begin{equation}
t(r,v) = t(0,v) + r \chi_{1}(v) + \frac{r^2}{2 !} 
\chi_{2}(v)+{\mathcal{O}}(r^3) \label {Dt22}
\end{equation}
where 
\[ \chi_{1}(v) = \frac{dt}{dr} \Big{|}_{r=0}, \; \; 
\chi_{2}(v) =  \frac{d^{2}t}{dr^{2}} \Big{|}_{r=0} .\]
Using equation (\ref{ARDef1}) in equation (\ref{Dsi21}), we obtain
\begin{equation}
 t(0,v) = t_{i} + \int_{v}^{1}  \frac{ e^{- \nu} dv}
 { \sqrt{\frac{M(0,v)}{v} +2 A_{1}(v) } } .
 \label {Dsi21A}
\end{equation}
 
The time for other collapsing shells to arrive at 
the singularity can be expressed by
\begin{equation}
 t_{s}(r)=t(r,0) = t_{i} + \int_{0}^{1}  \frac{ e^{- \nu} dv}
  {\sqrt{{\frac{M(r,v)}{v}} +\frac{ e^{2rA} -1}{r^2} }}.
\label {Dt26}
\end{equation}
Since the energy density has no
cusps at the center, that means $M_{1} = 0$. The singularity 
curve then takes the form,
\begin{equation}
t_{s}(r) = t_{0} + r \chi_{1}(0) + \frac{r^{2}}{2 !} 
\chi_{2}(0) + {\mathcal{O}}(r^{3}), \label{Dts26a}
\end{equation}
where $t_{0} = t(0, 0)$ is the time at which the central 
shell becomes singular. 
%%%%%%%%%%%%%%%%%%
Noting that $\nu \; '(0,v)=0$, we have
 \begin{equation}
 \chi_{1}(v)  = -  \int_{v}^{1} 
\frac{ e^{-\nu(0,v)} A_{2}(v)} {\left[\frac{M(0,v)} {v} 
+ 2 A_{1}(v) \right]^{3/2}} dv \label{CHI1}
\end{equation}
for all $k \in (0,1)$. The case $k = 0$ corresponds with the usual Lamaitre-Tolman-Bondi (LTB) dust solution
 which is discussed in section (V.1). If $f '''(0)=0$, then we have $\chi_{1}(0)= 0$ for all $k$,
 so we need to compute $\chi_{2}(0)$ for an 
analysis of the nature of the central singularity
in terms of its visibility or otherwise. We obtain, 
%%%%%%%%%%%%%%%%%%%%
\begin{eqnarray}
\chi_{2}(v)   = -     \int_{v}^{1}  e^{- \nu(0,v)}
  \Big\{ 	\frac{5 \; l}{3} \frac{e^{-2 Z(v)} }  
{ \left[\frac{M_{0}(v)} {v} + 2 A_{1}  \right]^{1/2}} \hspace{0.7in} \nonumber \\
 - \frac{3  A_{2}^2 } { \left[\frac{M_{0}(v)} {v} + 2 A_{1}  \right]^{5/2}} 
   +  \frac{ [\frac{ M_{2}}{v} + 2 A_{3} + 2 A_{1}^{2}] } 
{\left[\frac{M_{0}(v)} {v} + 2 A_{1}  \right]^{3/2}}
   \Big\} dv  \hspace{0.3in}\label{ch200A}
 \end{eqnarray}
%%%%%%%%%%%%%%%%%%%%% 
Since the term $[M_{0}(v)/v + 2 A_{1}(v)] > 0$, and with the 
suitable continuity of the functions involved in the integrand of 
equation (\ref{ch200A}), the sign of the term $\chi_{2}(v)$ is a resultant
of behavior of the terms $M_{0}, M_{2}, A_{1}, A_{2}$ and $ A_{3}$. 
Therefore the positive or negative sign of $\chi_{2}(0)$, for 
various values of $k \in (0,1)$, depends on the behavior of the 
functions $M_{0}, M_{2}, A_{1}, A_{2}$ and $ A_{3}$. 
We also note that the positive constant $m_{o}$ 
appearing in the mass function $F(t,r)$ is related with density 
by the expression $ 3 m_{o}= \rho(0,1)$. So, 
$m_{o}$ is in fact represents one-third of density 
of the star at the center taken on a cross section of an 
initial surface $v=1$ (at the beginning of the collapse at 
the instant $t=t_{i}$).
%%%%%%%%%%%%%%%
\subsection* {III.1 \ Special case when $g(v)=0$ at $r=0$}
In general $g(v)$ need not be zero at $r=0$ but to analyze the out come of end state 
of collapse as either a NS/BH phase, we are considering this 
particular case where $g(v)=0$ at $r=0$ i.e.
$g(v)$ vanishes at the central shell 
at all regular points of the spacetime.

Under this condition $Z= k \ln(v)$ and $e^{-Z}=v^{-k}$, 
and therefore  energy density at the center is expressed by
\begin{equation}
\rho(0,v) = \frac{3 \; m_{o} f(0)} {v^{3(k+1)} }. \label{CDEN}
\end{equation}
Next the first terms $A_{i}(v)$ in equation (\ref{ARDef1}) are obtained as
\begin{eqnarray}
 A_{1}(v) &&=  \frac{5 [v^{-2k} - 1]}{6(k+1)},  \nonumber \\
 A_{2}(v) &&=  \frac{f ''' (0) [1 -v^{-3k} ]} {3(k+1)} ,   \nonumber \\
 A_{3}(v) && =  \frac{( 7 f ''''(0) -25) } {72(k+1)}  [1 -v^{-4k} ]. \label{As17}
\end{eqnarray}

The center of the cloud is regular at all the epochs,
and time of reaching to such a regular center is expressed
using equation (\ref{Dsi21A}), 
\begin{eqnarray}
t(0,v) = t_{i} + (3 m_{o})^{l} 
  \int_{v}^{1}  \frac{ dv}
 { v^{3 k} \sqrt{\frac{M(0,v)}{v}  + 2 A_{1}(v) } }. \label{TR0}
\end{eqnarray}
The time at which the central 
shell becomes singular is obtained to the first order approximation,  
\begin {eqnarray}
 t(0,0)&&= t_{i} + 3^{l} (m_{o})^{l-3/2} \times 
\nonumber \\
&& \left\{\frac{2 \; m_{o}} {3(1-k)}  
 - \frac{5}{3(k+1)}\left[ \frac{1}{5-k} - \frac{1}{5+3k} \right] 
\right\}.  \label{Trv00}
\end{eqnarray}
The time taken by the central shell to reach the 
singularity should be positive and finite, and hence we have 
the {\it model realistic condition} (MRC), namely that,
\begin{equation}
\left\{\frac{2 \; m_{o}} {3(1-k)}  
 - \frac{5}{3(k+1)}\left[ \frac{1}{5-k} - \frac{1}{5+3k}\right] 
\right\} > 0 \label{MRC01}
\end{equation}
and it must be finite for any 
$k \in [0,1)$. The feasible region of MRC is shown in Fig. 1. 
%%%%%%%%%%%%%%%%%
\begin{figure}[ht]
\label{fig-01}
\begin{center}
\resizebox{7.0cm}{7.0cm}
{\includegraphics{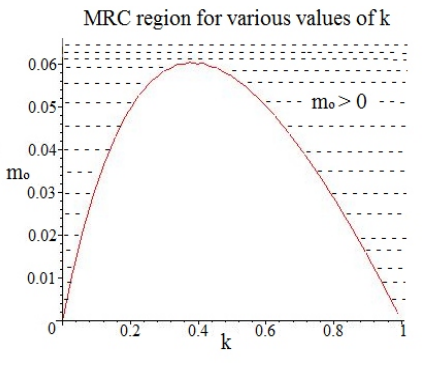}}
\caption{We have plotted the inequation (\ref{MRC01}).  The dotted lines region
shows the model realistic choices of $m_{o}$. 
There are ample choices of $m_{o}$, one of which fulfills its role
 as one-third of central density of the collapsing star at the initial epoch.
$m_{o} = 0.06$ approximately, being its l.u.b. for the set of all $k$,  $k \in [0,1)$. }
\end{center}
\end{figure}
%%%%%%%%%%%%%%%%%

The local increasing or decreasing nature of the singularity curve can be decided through the quantities
$\chi_{1}(0)$ and $\chi_{2}(0)$ which are computed as follows:
\begin{widetext}
\begin{eqnarray}
 \chi_{1}(0)  = 4 \; l\; 3^l m_{o}^{(l-3/2)}   f '''(0) \Big[ \frac{1}{(5+3k)(5-3k)} 
 - \frac{20 \; l (7+4k)}{m_{o} (7+9k)(7+5k)(7+3k)(7-k)} \Big]
\label{CHI10}
\end{eqnarray}
%%%%%%%%%%%
\begin{eqnarray}
 \chi_{2}(v) = - (3 m_{o})^{l}  
  \int_{v}^{1}  
  \left\{ \frac{5 \; l}{3} \frac{v^{-5k} } 
{ \left[\frac{M_{0}(v)} {v} + 2 A_{1}  \right]^{1/2}} 
- \frac{3 v^{-3k} A_{2}^2 } { \left[\frac{M_{0}(v)} {v} + 2 A_{1}  \right]^{5/2}} 
   +  \frac{ [\frac{ M_{2}}{v} + 2 A_{3} + 2 A_{1}^{2}] \; v^{-3k}} 
{\left[\frac{M_{0}(v)} {v} + 2 A_{1}  \right]^{3/2}}
   \right\} dv  \label{ch201}
 \end{eqnarray}
It then follows that,
{\small{
\begin{eqnarray}
  \chi_{2}(0)=  \frac{3^l m_{o}^{l-5/2}}{ 72  (k+1)^3}  
  \Big\{ \left[24 m_{o}^2  C_{0} + 16 k  m_{o} C_{1} + 250 k C_{2} \right]
	+ k f '''(0)^2 \left[ -\frac{200 }{m_{o}} C_{3} + 48  C_{4} \right] 
	 + k f ''''(0) \left[ 28 m_{o}  C_{5} - 70   C_{6} \right]	
	\Big\} 
		\label{ch202}
\end{eqnarray}
} }
  for all  $k \in (0,1)$ and where $C_{0}=(k+1)^2$,
 $C_{1} = \frac{ (13 k^3 - 129 k^2 -237 k -95)} { (1-k) (5+3k) (5-k)}$,
 $C_{2}=\frac{ 64 k (1+3k)} {(7+5k) (7-3k) (7+9k)}$, $C_{5}= \frac{8 (k+1)^2 } {5 (1-k) (5+3k)} $  \\
$C_{3}=  \frac{32 k^2 (115 k^2 + 378 k + 243)} {9 (k+1) ( 9 + 11k) (3 + k) (3 + 5k) (9+5k)(9-k)}$,
$C_{4}= \frac{72 k (k+1)} {(7+9k) (7+3k) (7-3k)}$,
and $C_{6}= \frac{64 k (7+3k) (k+1)} {(7+5k) (7+9k) (7+k) (7-3k)}$
are functions of $k$.
\end{widetext}
%%%%%%%%%%%%%%%%%%%%% 

\section {Radial null geodesics}
Now, we investigate below when there will be families of null geodesics
emanating, which will be future directed and outgoing, and which terminate 
in the past at the singularity, thus making
the communication from the singularity to an outside observer possible, 
as opposed to a black hole situation. Hence, for examination of the nature of
central singularity at $ R = 0, r = 0$, 
we consider the equation of outgoing radial null geodesics,  given by,
\[ \frac{dt} {dr} = e^{\psi - \nu}. \]
  Further, we write the null geodesic
equation in terms of the variables ( $ u= r^{\beta}, R $),
choosing $ \beta = \frac{1}{1-k}[\frac{5}{3} -k ]$ for $k \in (0,1)$, 
and using equation (\ref {Dsfe08}), we obtain
\begin{equation}
  \frac{dR}{du} = \frac{1}{\beta} \left( \frac{R}{u} +
\frac{ v'v^{\frac{1 -3k}{2}} } { (\frac{R}{u})^{\frac{[1 -3k]}{2}} } \right) \left( \frac{1 -
\frac{F}{R} } { \sqrt{G}[ \sqrt{G} + \sqrt{H}]} \right)  . \label
{rn25}
\end{equation}
If the null geodesics terminate at the singularity in the past
with a definite tangent, then at the singularity, we have 
$dR/du > 0 $, in the $(u, R)$ plane with a finite value.
For the case, when $R' > 0$ near the central singularity,
we have
\begin{equation}
x_{o} = \lim_{t\rightarrow t_{s}} \lim_{ r\rightarrow {o}}
\frac{R}{u} =  \frac{dR}{du} \Big{|} _{t\rightarrow
t_{s},r \rightarrow {o}} .\label {pr26}
\end{equation}
Using  equation (\ref{RD19}) for $v'$ and equation (\ref {rn25}), we determine
  \begin{eqnarray*}
  {x_{o}}^{\frac{3(1-k)}{2}} = \frac{3(1-k)}{2} \sqrt{m_{o}}
 \left(\frac{1} {3 m_{o}} \right)^{\frac{k} {k+1} }  \chi_{1}(0) 
 \end{eqnarray*}
 for $k \in (0,1)$. The radial null geodesic emerging from the
singularity in $(R,u)$ co-ordinates is $ R = x_{o} u$, or in
$(t,r)$ plane, it is given by 
\[ t - t_{s}(0) = x_{o} \; r^{\beta} \; . \]
Therefore, $ x_{o} > 0$ iff $ \chi_{1}(0) > 0 $, and hence $\chi_{1}(0) > 0$ 
is a sufficient condition for the occurrence of the NS at the center of the cloud 
as the end state of gravitational collapse of a sufficiently dense star
 when it loses its equilibrium state, and continual collapse begins 
with regular initial data of density and pressure profiles. 
This aspect is studied by many authors
 \cite{jg,dj2, rgj,rgpj-1}, 
there will be 
radially future outgoing null geodesics emanating from the singularity,
giving rise to a locally NS at the center. However, if $ \chi_{1}(0) < 0 $ then
 we have a black hole solution, as there will be no
such trajectories coming out. 

Thus the final fate of collapse of a massive star 
leads to either a BH or a NS with appropriate choice of initial data of 
 mass function through $m_{o}$, $f$ and the parameter of equation of state $k$. This fact is 
further illustrated  through the various figures. Fig.-2 exhibits dominant role played by $f '''(0)$
in deciding the sign of $\chi_{1}(0)$. Fig.-3  reveals the importance of initial central density $m_{o}$
and $f '''(0)$ in bringing about a change in sign of $\chi_{1}(0)$ as $m_{o} \rightarrow 1$.
Is there ample data in deciding the nature of singularity? This is being satisfactory answered through
the illustration given in Figs.-4 \& 7. What if  the central density is allowed to be kept fixed through 
$m_{o}=1$ unit but having varying equation of state, this question is answered
 through the illustration given in Figs.-5 \& 6.
 
\begin{figure}[ht]
\label{fig-02}
\begin{center}
\resizebox{7.0cm}{7.0cm}
{\includegraphics{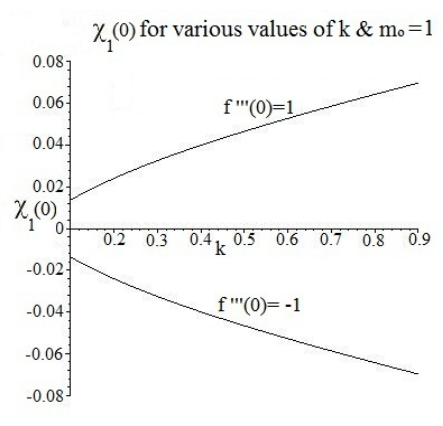}}
\caption{The nature of $\chi_{1}(0)$ is illustrated through the above
graphs. The condition $f '''(0)=1$ leads to formation of NS whereas $f '''(0) = -1$ confirms formation of BH.}
\end{center}
\end{figure}
%%%%%%%%%%%%%%%%%%%
\begin{figure}[ht]
\label{fig-03}
\begin{center}
\resizebox{7.0cm}{7.0cm}
{\includegraphics{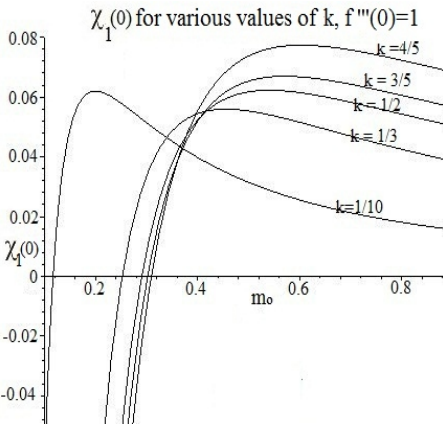}}
\caption{The above illustration of $\chi_{1}(0)$ indicates the role played by initial central density  through the 
mass the function. $f '''(0) =1$ together with higher initial central density through $m_{o}$
 propels formation of NS for all $k$. 
 }
\end{center}
\end{figure}
%%%%%%%%%%%%%%%%%%%%%%%%%%%%%%
%%%%%%%%%%%%%%%%%%%
\begin{figure}[ht]
\label{fig-04}
\begin{center}
\resizebox{7.0cm}{7.0cm}
{\includegraphics{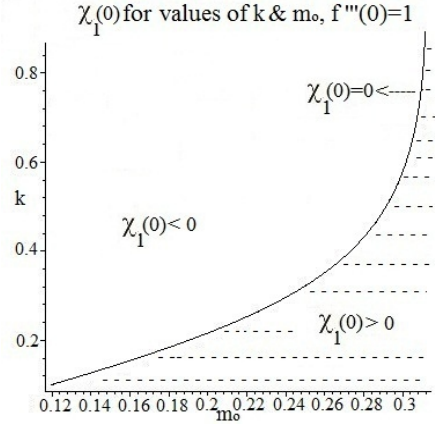}}
\caption{The above implicit graph illustrates the wide availability of initial data for the formation of 
BH and NS for various values of $k$ and $m_{o}$ under the condition $f '''(0)=1$. When
$f '''(0) =-1$ the regions of initial data for BH and NS are swapped.}
\end{center}
\end{figure}
%%%%%%%%%%%%%%%%%%%%%%%%%%%%%%
%%%%%%%%%%%%
\begin{figure}[ht]
\label{fig-05}
\begin{center}
\resizebox{7.0cm}{7.0cm}
{\includegraphics{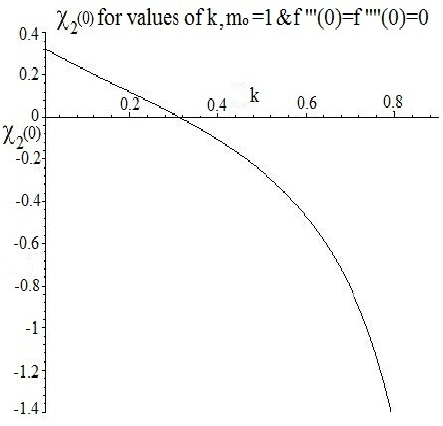}}
\caption{For a fixed value of the central density 
corresponding to $m_{o}=1$, the quantity 
$\chi_{2}(0)$ takes positive values for small positive values of $k$.
 Then it crosses the $k$-axis 
and takes negative values as $k \rightarrow 1$. 
The same graph can be obtained under other conditions $f ''''(0)= \pm 1$}
\end{center}
\end{figure}
%%%%%%%%%%%%%%%
%%%%%%%%%%%%%
\begin{figure}[ht]
\label{fig-06}
\begin{center}
\resizebox{7.0cm}{7.0cm}
{\includegraphics{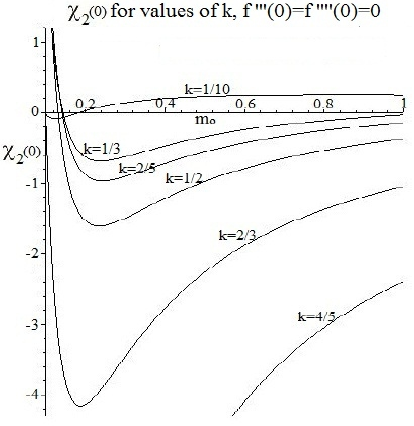}}
\caption{ We note that if the model begins with a higher value 
of $m_{o}$, higher values of $k$ would satisfy the condition 
$\chi_{2}(0) > 0$. Thus, it is the initial central density that 
can specify and would have ramification on the nature of 
singularity in terms of its visibility or otherwise. Almost the same graph can be obtained for the condition
$ f '''(0)= 0$ and $f ''''(0)= -1$.}
\end{center}
\end{figure}
%%%%%%%%%%%%%%%%%%%
\begin{figure}[ht]
\label{fig-07}
\begin{center}
\resizebox{7.00cm}{7.0cm}
{\includegraphics{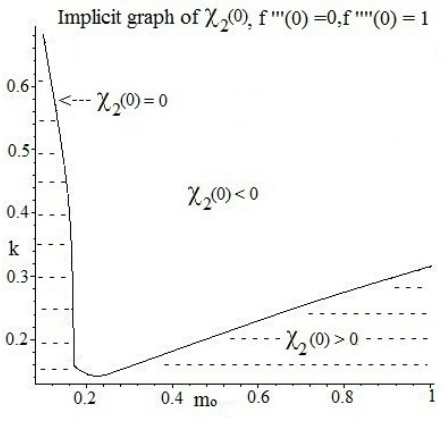}}
\caption{An implicit graph of $\chi_{2}(0)$ specifies the 
regions for formation of NS and BH,
as determined by the variation in parameters 
$m_{o}$ and $k$.
 }
\end{center}
\end{figure}
%%%%%%%%%%%%%%%%%%%%%%%%%%%%%%%%%%%
\subsection*{IV.1 The Dust collapse}
The LTB dust model is the most studied model in gravitational collapse. 
In view of the initiation of the of equation (\ref{CON01}), the detailed calculations of the dust case 
are presented in Appendix II.  It is observed that the introduction of the ansatz in equation (\ref{CON01}) 
restricted the class of dust solutions only to a homogeneous density profiles mainly because in dust case
$g(v) = 0$ for all $r$. Certainly, in such 
density profiles the final state of collapse leads to formation of black hole only. 

\subsection*{IV.2  The Radiation collapse and other values of $k$}
The radiation collapse can be considered through 
the case $k=1/3$ for which the
MRC takes the form 
 $m_{o}> 5/84$ and we have
\begin{eqnarray}
\chi_{{1}_{k=1/3}}^{g=0}(0)
 && =  \frac{ 3^{1/4} \; f '''(0)  (208 \; m_{o} - 45)} {4992 \; m_{o}^{9/4} } \label{CHI111} \\
 \chi_{{2}_{k=1/3}}^{g=0} (0) = 
 && \frac{3^{5/4}} {512 \; m_{o}^{9/4}}
  \Big[ \frac{128}{3}  \; m_{o}^{2} - \frac{10144}{189} m_{o}
    + \frac{800}{177} \nonumber \\
	&&	+ \frac{1}{3} f '''(0)^2 + \frac{1}{3} f ''''(0) \Big] . \label{CHI1302}
\end{eqnarray}
Now in equation (\ref{CHI111}) if  (i) $ f '''(0) > 0$, \; $m_{o} > 45/208$ or
(ii) $ f '''(0) < 0$, \; $5/84 < m_{o} < 45/208$, then in these situations we have $\chi_{{1}_{k=1/3}}^{g=0}(0) > 0$, 
otherwise, it takes negative value.

Next in equation (\ref{CHI1302}), the bracketed expression has varying sign, 
we find that $\chi_{{2}_{k=1/3}}^{g=0} (0) > 0$ for small positive values 
of $m_{o}$, and for higher values of $m_{o}$, it changes its sign.

In general, to understand the nature of $\chi_{i}(0)$ relative to $k$, 
we can obtain expressions for $\chi_{i}(0)$ for 
different values of $k$ in order to see how it
changes with a change in the values of parameter $k$. This is summarized in figs. 3 \& 6.

%%%%%%%%%%%%%%%%%%%%%%%%%%%%%%
\section {The apparent horizon}
Formation of trapped surfaces is fundamental in the study of end state of gravitational collapse
and apparent horizon is the boundary of the trapped surfaces. 
So when the singularity curve is increasing, 
it could be shown that the apparent horizon which begins also at the 
central singularity at $r=0$, is increasing necessarily, and 
therefore it follows that there is no formation of trapped 
surfaces as the gravitational collapse of the matter cloud evolves, 
all the way till the singularity formation epoch. In such a
case, as has been shown earlier
\cite{dj2},
the singularity is necessarily locally naked and a family
of future directed null (and also timelike) geodesics come out
which terminate in the past at the singularity. Such a singularity
can also be globally visible depending on the nature of the
mass function away from the center
\cite{dj2}.    
On the other hand, when the singularity curve is
constant ($\chi_i =0$),
or would be decreasing, then a black hole will necessarily
form as the collapse final state.

The apparent horizon is the boundary of trapped surfaces which in general is
given by the equation
\cite{DPJRVS}
\begin{equation}
1 - \frac{F} {R} =0  \; .\label{APP01}
\end{equation} 
Using values of $F$ and $R$, the equation of the curve $v_{ah}(r)$ is obtained as
\begin{equation}
 v_{ah}(r) = r^2 M(r, v_{ah}) = m_{o} r^2 f( r  e^{-  Z(v_{ah})})  e^{- 3 Z(v_{ah})}, \label{APP02}
\end{equation} 
and to note that $v_{ah}(r)$ is a positive real root of the above equation.

Inversely, the $t_{ah}(r)$ curve represents the apparent horizon curve that 
gives the time at which the shell labeled by $r$ becomes trapped.
We can evaluate the time curve $t_{ah}(r)$  of the apparent horizon to determine the
visibility of the singularity to the external observer. We have
\begin{equation}
 t_{ah}(r) = t_{s}(r) - \int_{0}^{v_{ah}(r)}  \frac{ e^{- \nu} dv}
 { \sqrt{\frac{M(r,v)}{v} +\frac{ e^{2rA} -1}{r^2} }}.
 \label {APP03}
\end{equation}
Now using expression for $t_{s}(r)$ given by equation (\ref{Dt26}),
  above equation takes the form
\begin{eqnarray}
  t_{ah}(r) = t_{i} + \int_{v_{ah}(r)}^{1}  \frac{ e^{- \nu} dv}
 { \sqrt{\frac{M(r,v)}{v} +\frac{ e^{2rA} -1}{r^2} }} . 
 \label {APP03a}
\end{eqnarray}

We have basically, a naked singularity could occur when a comoving 
observer (at a fixed radius $r$) does not encounter any 
trapped surfaces until the time of singularity formation. 
On the other hand, for a black hole the trapped surfaces
form before the singularity. Thus for a black hole to
form we require,
\begin{equation}\label{bh}
t_{\rm ah}(r)\leq t_0~\mbox{for}~r>0\,,~\mbox{near}~r=0\,
\end{equation}
where $t_0$ is the epoch at which the central shell
hits the singularity.

Near $r =0$, equation (\ref{APP03a}) can be written as
\begin{equation}
t_{ah}(r) = t_{0} + r \; \chi_{1}(v_{ah}) + \frac{r^{2}}{2 !} 
\chi_{2}(v_{ah}) + {\mathcal{O}}(r^{3}). \label{APP04}
\end{equation}
In general, condition (\ref{bh}) is violated when
$\chi_1(v_{ah}) > 0$ or $\chi_2(v_{ah}) > 0$ ( if $\chi_1(v_{ah})=0$ )
as seen from the above equation.
The apparent horizon curve that initiates at the singularity 
$r=0$ at the epoch $t_0$, then increases with increasing $r$,
 moving to the future and we have 
$t_{\rm ah} > t_0$ for $r > 0$ near the center. The behavior of 
the outgoing families of null geodesics has been analyzed in detail 
in these cases, and it is known that the geodesics terminate 
at the singularity in the past, which results in a naked singularity. 
In such cases then the extreme strong gravity regions can 
communicate with external observers. 
\cite{RGPJ} 
We have
 \begin{eqnarray}
 \chi_{1}(v_{ah})  = -  \int_{v_{ah}}^{1} 
\frac{ e^{-\nu(0,v)} A_{2}(v)} {\left[\frac{M(0,v)} {v} 
+ 2 A_{1}(v) \right]^{3/2}} dv \label{CHI11} . 
\end{eqnarray}
Here, it is a must that $[M(0,v)/v+ 2  A_{1}(v) ] > 0$ and that the sign of $A_{2}(v)$ has become all that important.
So if $A_{2}(v) < 0$ then the integrand of  $\chi_{1}(v_{ah})$ becomes a positive continuous function,
and therefore $\chi_{1}(v_{ah}) > 0$, giving rise to $t_{ah} > t_{o} $ for some $r$ near the center.
This increasing apparent horizon curve of light rays or particles emanating
 from the singularity $r=0$ will be at least locally visible to the observer,
 confirming the formation of at least locally NS.

 The other condition that $A_{2}(v) > 0$ emphasizes
formation of BH with $\chi_{1}(v_{ah}) < 0$. 

 Also, we observe that $A_{2}(v) =0 \Leftrightarrow f ''' (0)=0$,
and in this case $\chi_{1}(v_{ah}) = 0$, therefore the analysis of apparent horizon can be carried out through
$\chi_{2}(v_{ah})$ ; 
\begin{eqnarray}
 \chi_{2}&&(v_{ah}) = -     \int_{v_{ah}}^{1} \frac{ e^{- \nu(0,v)}} {{\left[\frac{M_{0}(v)} {v} + 2 A_{1}  \right]^{3/2}}}
  \Big\{ 	\frac{5 \; l}{3} e^{-2 Z(v)}  \nonumber \\
	&& \times \left[\frac{M_{0}(v)} {v} + 2 A_{1}  \right]
   +   \left[\frac{ M_{2}}{v} + 2 A_{3} + 2 A_{1}^{2} \right]  
   \Big\} dv.  \label{ch200B}
 \end{eqnarray}
In the above expression, the first term in curly bracket is always positive, 
so the sign of the integrand is mainly focused on the nature of $M_2$ and $A_3$. 
If the values of these terms are sufficiently negative, in that case sign of the integrand can be positive,
and thereby $\chi_{2}(v_{ah}) > 0$, otherwise we have $\chi_{2}(v_{ah}) < 0$.
Thus the nature of increasing or decreasing apparent horizon depends on the initial data of
mass function through $M(r,v)$ and metric function $\nu$ associated with the density profiles.   \\

In the {\it particular case} when we use $g(v)=0$ at $r=0$ in $\chi_{1}(v_{ah})$, we have
\begin{eqnarray}
\chi_{1}^{g=0}(v_{ah})&&= - \frac{[3 m_{o} f(0)]^l f'''(0)} {3(k+1)}   \nonumber \\
&& \times \int_{v_{ah}}^{1}  
\frac{  v^{-3k+3/2}   ( 1 - v^{-3k})} {\left[M(0,v) 
+ 2 v A_{1}(v) \right]^{3/2}} dv.  \label{CHI14} 
\end{eqnarray}
On using values of $M(0,v)$ and 
$A_{1}(v)$, and solving $[M(0,v)+ 2 v A_{1}(v) ]$,
 we obtain $ v^{k+1} (1- v^{2k}) > -3/5  \ m_{o} (k+1)$, and this condition holds good
since $ v^{2k} \leq 1$ for all $k \in (0,1]$.

In the integrand of equation (\ref{CHI14}), we cannot have
 the term $( 1- v^{-3k}) > 0 $, since this implies $v^{3k} > 1$
 and this is not possible as $v \in [0,1]$. So, there are only two conditions under which sign of 
$\chi_{1}^{g=0}(v_{ah})$ can be decided, namely \\
(i)$ f '''(0)  > 0$ and $( 1- v^{-3k}) < 0 $, \\
 (ii) $f '''(0) < 0$ and $( 1- v^{-3k}) < 0 $. \\
Under condition (i),  $\chi_{1}(v_{ah}) > 0$, implying $t_{ah} > t_{o} $ for some $r$, near $r =0$,
revealing increasing nature of apparent horizon curve which
 confirms formation of at least locally NS, while under condition (ii) the 
BH forms with $\chi_{1}^{g=0}(v_{ah}) < 0$.
These results are in agreement with the results shown in fig. (2) using sufficient condition for 
formation of NS/BH phases.

We observe that $\chi_{1}^{g=0}(v_{ah}) = 0 \Leftrightarrow f ''' (0) =0$ and when $\chi_{1}^{g=0}(v_{ah}) =0$,
 the study of apparent horizon curve is studied through  $\chi_{2}^{g=0}(v_{ah})$. 
\begin{widetext}
\begin{eqnarray} 
\chi_{2}^{g=0}(v_{ah})= - (3 m_{o})^{l}  
  \int_{v_{ah}}^{1}  && \Big\{ \frac{v^{3/2}} {\left[M_{0}(v)  + 2 v A_{1}  \right]^{3/2}}
	\Big[	\frac{5 \; l}{3} v^{-5k}  \left( m_{o} v^{-3k-1} + \frac{5 (v^{-2k} -1) } {3 (k+1)} \right) \nonumber \\
	&& + \left(- \frac{ m_{o} v^{-5k-1}} {2}  + \frac{2 (7 f ''''(0) -25) (1 - v^{-4k})} {72 (k+1)}
	+ \frac{25 (v^{-2k} -1)^2 } {18 (k+1)^2} \right) \; v^{-3k} \Big]
    \Big\} dv  \label{ch23} 
\end{eqnarray} 
 \text{Also we find}  $\chi_{2}^{g=0}(v_{ah})$ \text{for a particular value of } $k =1/10$  \text{with} 
   $m_{o} =1$  \text{and}  $f ''''(0) =0$. 
\begin{eqnarray}
\chi_{{2}_{k=1/10}}^{g=0}(v_{ah})= - (3 m_{o})^l  
  \int_{v_{ah}}^{1} \Big\{ \frac{v^{3/2}} {\left[M_{0}(v)  + 2 \; v A_{1}  \right]^{3/2}}
	\left( \frac{  1125 \; v^{3/2} + 4375 \; v^{11/10} - 5500 \; v^{13/10} - 759  } {2178 \; v^{9/5} }\right)
    \Big\} dv  \label{ch24}  
\end{eqnarray} 
\end{widetext}
It is observed that in the above expression, the term in round bracket is negative in sign for $v \in (0,1]$.
Thus for $k=1/10$, $\chi_{{2}_{k=1/10}}^{g=0}(v_{ah}) > 0$, and therefore the apparent horizon curve is increasing 
in the neighbourhood of the central singularity, making it at least locally naked. Similar results can be obtained
for $k =1/100$, etc; reflecting the role played by pressure, the smallest addition of pressure
in inhomogeneous dust collapse can lead to formation of naked singularity as shown in the fig.8.
These results are in agreement with earlier conclusions drawn from $\chi_{1}(0)$ and $\chi_{2}(0)$.

Thus, what we observe is that
the increasing or decreasing  nature of apparent horizon curve solely depends on the choice of
 $m_{o}, k$, $f '''(0)$ and $f ''''(0)$.  
%%%%%%%%%%%%%%%%%
\begin{figure}[ht]
\label{fig-08}
\begin{center}
\resizebox{7.0cm}{7.0cm}
{\includegraphics{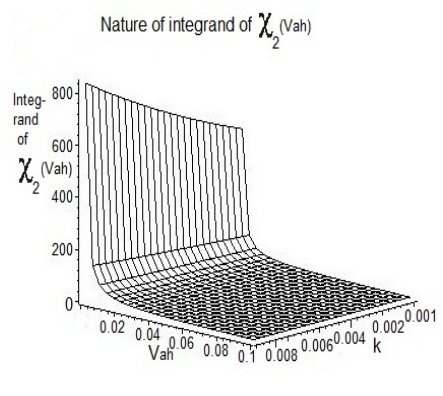}}
\caption{Through the positive nature of the integrand of $\chi_{2}(v_{ah})$, it is revealed that
even for small values of pressure through parameter $k$, the apparent horizon curve is increasing.  }
\end{center}
\end{figure}
%%%%%%%%%%%%%%%%%%

For the dust case, $k=0$, equation (\ref{APP03}) takes the form
\begin{equation}
 t_{ah}(r) = t_{s}(r) - \int_{0}^{v_{ah}(r)}  \frac{ \sqrt{v} dv}
 { \sqrt{m_{o} } \sqrt{f(r)} }. 
 \label {APP33}
\end{equation}
On solving above equation and in this case $t_{s}(r) = t_{0}$ (since $\chi_{i}(0) = 0$ ), we have
\begin{equation}
 t_{ah}(r) = t_{0} - \frac{2} {3 \sqrt{m_{o}} \sqrt{f(r)} } \; v_{ah}^{3/2}
 \label {APP34}
\end{equation}
and therefore  $t_{ah}(r) - t_{0} < 0$ as expected for the BH case explained in the Appendix II.

Now, we are mainly focused in the case where the central singularity would
be visible but in general, there can be situations where non-central singularity becomes visible.
This is possible in
the case in which $M(r, v)$ goes to zero as $v$ goes to zero, leaving $M/v$ bounded.  
\cite{PJM}
We probe this scenario in our case.

Using equations (\ref{Dsm12}) and (\ref{SoD15}) in equation (\ref{APP01}), we obtain
\[ r^{2} m_{o} f(x) e^{-3 Z(v)} = v \ . \]
It is clear that as $v \rightarrow 0$, we must have 
$r \rightarrow 0$ on the apparent horizon.  Certainly, it is not possible 
to satisfy $1 - F/R \ > 0$, {that is since $f(r)>0$}, we have
\[1- \frac{ r^{2} m_{o} f(x) e^{-3 Z(v)} } 
{v} < 0 \; \text{as}\; v \rightarrow 0 \; \text{for}\; r >0\]
near the singularity and away from center, that is 
with $r > 0$ or to say $M/v$ becomes unbounded  . It follows that the region surrounding the 
singularity cannot be timelike, and therefore any singularity 
that might eventually form near the center with $r > 0$ 
must not be visible. Therefore, the only 
singularity that can eventually be visible is that at the 
center of the collapsing cloud.

\section{Conclusions}
The investigation of gravitational collapse with a linear 
equation of state has revealed role of the parameter $k$ in deciding 
the nature of central singularity for the given initial data 
set of the mass function $M(r,v)$. We examined here the marginally
bound case with $b(r)=0$. 

In particular, we gave here an explicit class of mass 
functions that satisfies the Einstein equations and the metric 
obtained represents a class of solutions for a unique 
choice of functions $g(v)$, representing the exact solution 
for a perfect fluid sphere collapse with an isentropic 
equation of state. 

The occurrence of a locally naked singularity as collapse 
end state is shown for these collapse models for  
a wide variety of values of $k \in [0,1)$, and also formation of 
black hole is seen for a range of values of $k$  for this 
class of mass functions which are an appropriate choice of 
$M(r,v)$ that are solutions to the Einstein field equations. 
This is depicted through various figures which emphasizes 
the role played by parameter $k$ and initial central density
thorough $m_{o}$.

We thus see, through an explicit demonstration, that
both black holes and naked singularities arise naturally 
as collapse final states for perfect fluid collapse
with a linear isentropic equation of state. The class of
solutions we gave here have several intriguing properties
some of which we discussed here. 
It is thus seen that the choice of $k \in [0,1)$ 
influences and contributes to decide in the formation of 
NS/BH phases of gravitational collapse. The model 
realistic condition discussed here emphasizes the role 
of pressure in the gravitational collapse of the star.

We note that the present model can be extended for a 
more general case with $b(r) \neq 0$, and the results on the 
same shall be reported elsewhere.  

{\bf{Acknowledgement}}: \\
We like to thank Daniele Malafarina for his valuable comments. 
Sanjay Sarwe acknowledges the facilities extended by IUCAA, Pune, 
while part of this work was being completed.

%\begin {references}

%\end {references}

\section*{Appendix- I \\ Compatibility condition}
The introduction of the equation
(\ref{CON01}) which is a transformation between coordinates $(t,r)$ and $(r,v)$ given by 
$v'(t,r)= v g(v)/r$, demands its compatibility with other field equations or their subsequent 
equation (\ref{RD19}). So, we consider $v' = v g(v)/r = W(r,v)$ and $\dot{v} = U(r,v)$. The condition of compatibility for non-linear partial differential equations of order one yields,
\begin{equation}
W U_{,v} -  U W_{,v} = - U_{,r}. \label{AC-01}
\end{equation}
where $U(r,v) = - e^{\nu(r,v)} \sqrt{D(r,v)} $ and 
\begin{equation}
 D(r,v) = \frac{M(r,v)}{v} + \frac{e^{2rA(r,v)}-1} {r^2} \label{AC-02}
\end{equation}
Equation (\ref{AC-01}) takes the form
\begin{eqnarray}
-\frac{e^{\nu}}{ 2 r \sqrt{D}} \Big[ 2D \left( v g(v) \; \nu_{,v} -g(v) 
- v \frac{\partial{g}}{\partial{v}}
+ r \; \nu_{,r} \right) \nonumber \\
+ v g(v) D_{,v} + r D_{,r}\Big] =0 \hspace{0.6in} \label{AC-03}
\end{eqnarray}
Using equation (\ref{AC-02}), and values of $\nu_{,r}$ , $\nu_{,v}$ ( and subsequently
 $\rho$, $\rho_{,v}$ and $Z_{,v}$ ) above equation becomes
\begin{eqnarray}
&& 2 D [v(1 + g(v)) r A_{,v} - g(v)-v \frac{\partial{g}}{\partial{v}}]   \nonumber \\
 && +  \frac{2}{r^2}[r e^{2rA} (A + r A_{,r})- (e^{2rA}-1) ]  \nonumber \\
 && +  \frac{M_{o} \; x \; f'(x) e^{-3Z} }{v^2} [ v - \frac{3k}{y} ]  \nonumber \\
 && +  v g(v) \left[- \frac{M_{o} f(x) e^{-3Z} }{v^2} ( 1 + \frac{3k}{y} ) 
  +  \frac{2 e^{2rA} A_{,v} } {r} \right] \hspace{0.3in} \nonumber \\
 && +  \frac{2 k D}{(k+1) y} \Big\{ g(v)(k+1) [ 3(1+g(v)) + v \frac{\partial{g}}{\partial{v}} ]  \nonumber \\
 && +  \frac{ (1+g(v)) x [ 4 f'(x) + x f''(x)]} {[ 3 f(x) +x f'(x)]} \Big\}=0  \label{AC-04}
\end{eqnarray}

where $x= re^{-Z(v)}$ and $y=[1+(k+1)g(v)]$. Since $Z(v)$ and $A_{,r}$
are integration terms, therefore, above equation is an integro-differential equation.
 The dependent function $g(v)$ can be obtained as its solution in terms of $v$'s and integration constant for a chosen physically realistic function $f$. In turn, $v'$ will have a regulated form consistent with the field equations. So, $v'$ and $\dot{v}$ given by equations (\ref{CON01}) and
(\ref{RD19}) respectively, together determine $v$, and hence $R(t,r)$
 to obtain the requisite exact solution of the Einstein's field equations. \\
 
We consider a special case of $f(x) = 1- x^2 /2$ to understand the above scenario.
Equation (\ref{AC-04}) on using equation (\ref{ARDef1}) takes the form
\begin{eqnarray}
&& 2[E_{3} + E_{4}] \Big\{ - v \frac{dg}{ dv} -  g(v) + l v r^2  [ 1 + g(v) ] E_{5} 
 + \frac{2l} { (6-5 x^2) y }  \; \times\nonumber \\
&& \Big[ \frac{g(v)}{2} (6 -5 x^2) ( k+1) [ 3 (1 + g(v)) + v \frac{dg}{dv} ]
+ 5 r x( 1 +g(v))  \Big] \Big\} \nonumber \\
&& + g(v) \Big[ -\left( 1 + \frac{3}{y} \right) E_{4} + l v ( 2 + 4 r^2 E_{3}) E_{5} \Big] \nonumber \\
&& + 2 \Big[ ( 1+2r^2 E_{3}) \left( E_{3} + [\frac{5}{6}E_{1} + \frac{3 r^2}{72} E_{2} ] \right) -  E_{3} \Big] =0 
\label{CONf(x)}
\end{eqnarray}
where 
\begin{eqnarray*}
&& E_{1}= \frac{v^{-2k}-1}{k+1}, E_{2}= 25\frac{(v^{-4k}-1)} {k+1}, E_{3}= \frac{5}{6}E_{1} + \frac{r^2}{72}E_{2} \nonumber \\
&& E_{4} = \frac{m_{o} ( 1- x^2/2) e^{- 3 Z(v)}} {v} \; \text{and} \nonumber \\
 && E_{5} = - \left[ \frac{5 v^{-2k-1}}{3}  + \frac{25 r^2 v^{-4k-1}}{18}  \right].
\end{eqnarray*}
This is an integro-differential equation in $g(v)$, 
its solution satisfying initial conditions shall pave way for determining the exact solution of set of field equations.
Thus there is a certain class of functions available for f(x) that satisfy conditions on $f$ so as to have regular density and for such choice of $f(x)$, we will have some $g(v)$ 
that satisfies equation (\ref{AC-04}).

\section*{Appendix- II \\ The dust collapse}

In the well-studied case of the 
dust collapse, we have $p=0$ at $k=0$ and the MRC reduces 
to the form \ $m_{o} > 0$.  Herein, metric function $\nu =0$ and also the function $A(r,v)=0$. 
Therefore equation (\ref{RD19}) becomes
\begin{equation}
\dot{v} = -  \sqrt{\frac{M(r,v)}{v}  } . \label{DC01}
\end{equation}
On integrating, its solution at the surface $t=t_{s}$ is given by
\begin{equation}
v^{3/2} = \frac{3}{2} \sqrt{m_{o} } \sqrt{f(r)} ( t_{s} - t).   \label{DC02}
\end{equation}
Hence, equation (\ref{CON01}) takes the form
\begin{equation}
g(v) = \frac{2}{3} \frac{ r f'(r)} {f(r)}  \label{DC03} 
\end{equation}
but now $g$ can be a constant function only 
and for this, we have 
\begin{equation}
 \frac{ r f '(r)} {f(r)} = B_{1}  \label{DC04} 
\end{equation}
where $B_{1}$ is a constant.  Its solution is $f(r) = B_{2} r^{B_{1}}$ and in this case 
$M(r,v) = M(r) = {m_{o}} B_{2} r^{B_{1}}$.  The demand of the regularity condition that
the central density be positive and finite at the initial surface i.e.
$M(0) = {m_{o}} \neq 0$ and ${m_{o}}$ be finite, constraints $f(r)$ by imposing condition that
$B_{1}=0$.  In respect to this, the density profile takes the form
\begin{equation}
\rho=  \frac{ 3 \; B_{2}} { v(t)^3 } \label{DC05} 
\end{equation}
where $B_{2}$ is a positive constant. 
The initial data of density profiles is restricted to the homogeneous form only.
 \end{document}